# Longitudinal assessment of demographic representativeness in the Medical Imaging and Data Resource Center Open Data Commons

**Authors:** Heather M. Whitney[1,10*], Natalie Baughan[1,10], Kyle J. Myers[2,10], Karen Drukker[1,10], Judy Gichoya[3,10], Brad Bower[4,10], Weijie Chen[5,10], Nicholas Gruszauskas[1,10], Jayashree Kalpathy-Cramer[6,10], Sanmi Koyejo[7,10], Rui C. Sá[4,8,10], Berkman Sahiner[5,10], Zi Zhang[9,10], Maryellen L. Giger[1,10]

**Affiliations:** 1 University of Chicago, 2 Puente Solutions LLC, 3 Emory University, 4 National Institutes of Health, 5 United States Food and Drug Administration, 6 University of Colorado, 7 Stanford University, 8 University of California, San Diego, 9 Jefferson Health, 10 The Medical Imaging and Data Resource Center (midrc.org)

*Corresponding author
Heather Whitney
5841 S Maryland Ave., MC2026
Chicago, IL 60637
E-mail: hwhitney@uchicago.edu
Phone number: (773) 834-5097

**Abstract**

**Purpose:** The Medical Imaging and Data Resource Center (MIDRC) open data commons was launched to accelerate the development of artificial intelligence (AI) algorithms to help address the COVID-19 pandemic. The purpose of this study was to quantify longitudinal representativeness of the demographic characteristics of the primary imaging dataset compared to the United States general population (US Census) and COVID-19 positive case counts from the Centers for Disease Control and Prevention (CDC).

**Approach:** The Jensen Shannon distance (JSD) was used to longitudinally measure the similarity of the distribution of (1) all unique patients in the MIDRC data to the 2020 US Census and (2) all unique COVID-19 positive patients in the MIDRC data to the case counts reported by the CDC. The distributions were evaluated in the demographic categories of age at index, sex, race, ethnicity, and the intersection of race and ethnicity.

**Results:** Representativeness the MIDRC data by ethnicity and the intersection of race and ethnicity was impacted by the percentage of CDC case counts for which data in these categories is not reported. The distributions by sex and race have retained their level of representativeness over time.

**Conclusion:** The representativeness of the open medical imaging datasets in the curated public data commons at MIDRC has evolved over time as both the number of contributing institutions and overall number of subjects has grown. The use of metrics such as the JSD support measurement of representativeness, one step needed for fair and generalizable AI algorithm development.

## Introduction

Since the first identification of the SARS-CoV-2 coronavirus (and its associated infectious disease, COVID-19) in late 2019, there have been reports of differences in disease health outcomes by race, ethnicity, sex, and other demographics.[1–14] Additionally, the relative difference in impact of COVID-19 to demographic subgroups has also changed over time.[15,16] Furthermore, differences in the utilization of medical imaging in healthcare have been observed in various demographic subgroups over the course of the pandemic.[17–19] As a result, the use of medical imaging among different demographic subpopulations, for both COVID-19-related reasons or otherwise, can be expected to change over time.

The Medical Imaging and Data Resource Center (MIDRC) is a multi-institutional initiative designed to collect, curate, and share medical images and other related resources to support the development of artificial intelligence/machine learning (AI/ML) for diagnosis, treatment, and prognosis of COVID-19 and beyond. MIDRC is hosted at the University of Chicago, funded by the National Institute of Biomedical Imaging and Bioengineering, and co-led by the American College of Radiology® (ACR), the Radiological Society of North America (RSNA), and the American Association of Physicists in Medicine (AAPM). Studies are contributed to MIDRC by institutions via a pipeline that includes a collaborative partnership between the ACR®, the RSNA, the AAPM and Gen 3, a data commons organization. MIDRC places a strong emphasis on monitoring and increasing the representativeness of the data, both at specific instances in time and longitudinally, to help support the development of unbiased and generalizable algorithms.

The purpose of this study was to (1) introduce the use of a metric to measure representativeness of the imaging datasets compared to relevant groups and (2) report on the evolution of the representativeness since the ingestion of datasets from contributors began in August 2021.

## Materials and Methods

### *Dataset*

Data used in this study was composed of metadata for the imaging studies available at the MIDRC open data commons[20] in the Open-A1 and Open-R1 datasets (i.e., those ingested by the ACR® and RSNA and curated and harmonized by AAPM and Gen3). In this study, we refer to this specific collection as “MIDRC data.” The metadata had been submitted by data contributors in accordance with the MIDRC data dictionary.[21] Assignment of unique patients into the open data commons occurs at the ingestion of data within MIDRC according to a multidimensional stratified sampling algorithm,[22] with approximately 80% of unique patients assigned to the open data commons and 20% of unique patients assigned to a sequestered data commons. The sequestration algorithm is designed and tested for balance among groupings including but not limited to demographic categories.

For the purposes of this study, demographic categories were analyzed as follows: age at index event (i.e., the first occurrence in MIDRC, usually the first COVID-19 test), sex, race, ethnicity, and the intersection[23] of race and ethnicity (the latter in recognition that individuals belong to multiple groups). Patients may have multiple imaging studies in MIDRC but, for each unique patient, the characteristics at the index event were used in this study.

***Comparison groups***

The demographic distributions of unique patients in the MIDRC data were compared against two relevant population distributions. Because at the time of this study all Open-A1 and Open-R1 data contributing to the MIDRC data described here had been collected in the United States, we compared the demographic distributions (1) between all cases (unique patients) within the MIDRC data and population in the United States 2020 Census[24] and (2) between COVID-19 positive cases within the MIDRC data and population in the COVID-19 Case Surveillance Public Use Data from the Centers for Disease Control and Prevention (CDC).[25]

***Statistical analysis***

The Jensen-Shannon distance[26–28] (JSD) was used as a metric to measure the difference between any two population categorical distributions, with the two comparison groups being termed *S* and *T* in this study. It is based upon the Jensen-Shannon divergence[29] (called $D_{JS}$ in this study) and the Kullback-Leiber divergence $D_{KL}$. The $D_{KL}$ is defined for two distributions as

$$D_{KL}(S||T) = \sum_x S(x) log_2 \frac{S(x)}{T(x)} \quad (1)$$

where *S(x)* and *T(x)* are the distribution functions of any two populations *S* and *T*, and *x* is the variable of interest which in this study is any of the demographic variables under investigation. Because all the demographic variables *x* in this study are discrete, the distribution function (*S(x)* and *T(x)*) is a probability mass function (i.e., represented by the fraction of patients at each bin of *x*). Subsequently, the JSD is defined as

$$\text{JSD} = \sqrt{D_{JS}(S||T)} \quad (2)$$

where

$$D_{JS}(S||T) = \frac{1}{2} D_{KL}(S||M) + \frac{1}{2} D_{KL}(T||M) , \quad (3)$$

and

$$M = \frac{1}{2}(S + T). \quad (4)$$

The logarithm within $D_{KL}$ can be determined through other bases (such as the natural logarithm). When $\log_2$ is used, the $D_{JS}$ and JSD are bounded between 0 and 1, which is advantageous for our purpose. The sum in equation (1) was taken over each bin $x$ for which both comparison groups were non-zero. A JSD of zero indicates that there is no difference between compared distributions, while a JSD of 1 indicates that there is no similarity between them. In this study, more representative distributions (compared to the reference distribution) will have a lower JSD.

In this study, the JSD was used to compare the following distributions at each MIDRC batch ingestion date:

(1) cumulative counts of all unique patients in the MIDRC data to the US Census counts ($JSD_{\text{MIDRC (all) to Census}}$),

(2) the cumulative counts of all unique COVID-19 positive cases in the MIDRC data to the cumulative COVID-19 positive counts (derived from case counts) reported by the CDC ($JSD_{\text{MIDRC (C19+) to CDC (C19+)}}$), and

(3) the cumulative COVID-19 positive counts reported by the CDC to the US Census counts ($JSD_{\text{CDC (C19+) to Census}}$).

The CDC to US Census comparison was used as a reference against which the comparison of MIDRC distributions can be considered.

Additionally, the temporal difference in the JSD was determined between the JSD when comparing all cases in MIDRC to the US Census and the JSD when comparing all COVID-19 positive cases to the COVID-19 positive case counts from the CDC. If this difference is positive, the distribution of COVID-19 positive unique patients in MIDRC to the CDC cumulative COVID-19 case counts is more representative than the distribution of all unique patients in MIDRC to the US general population. If this difference is negative, the distribution of all unique patients in MIDRC to the US general population is more representative than the distribution of COVID-19 positive unique patients in MIDRC to the CDC cumulative case counts.

Note that in this study, no measures of statistical difference were assessed, since the goal of the measure of representativeness here is to measure degree of similarity according to counts. Additionally, no sampling of distributions was conducted, due to the nature of the data (counts of individuals), none of which are inherently considered samples in this study.

## Results

### *Dataset*

As of August 19, 2022 (the most recent batch ingestion date at time of manuscript preparation), there were 6 unique contributing sites and over 40,000 unique patients represented in the MIDRC data (Figure 1).

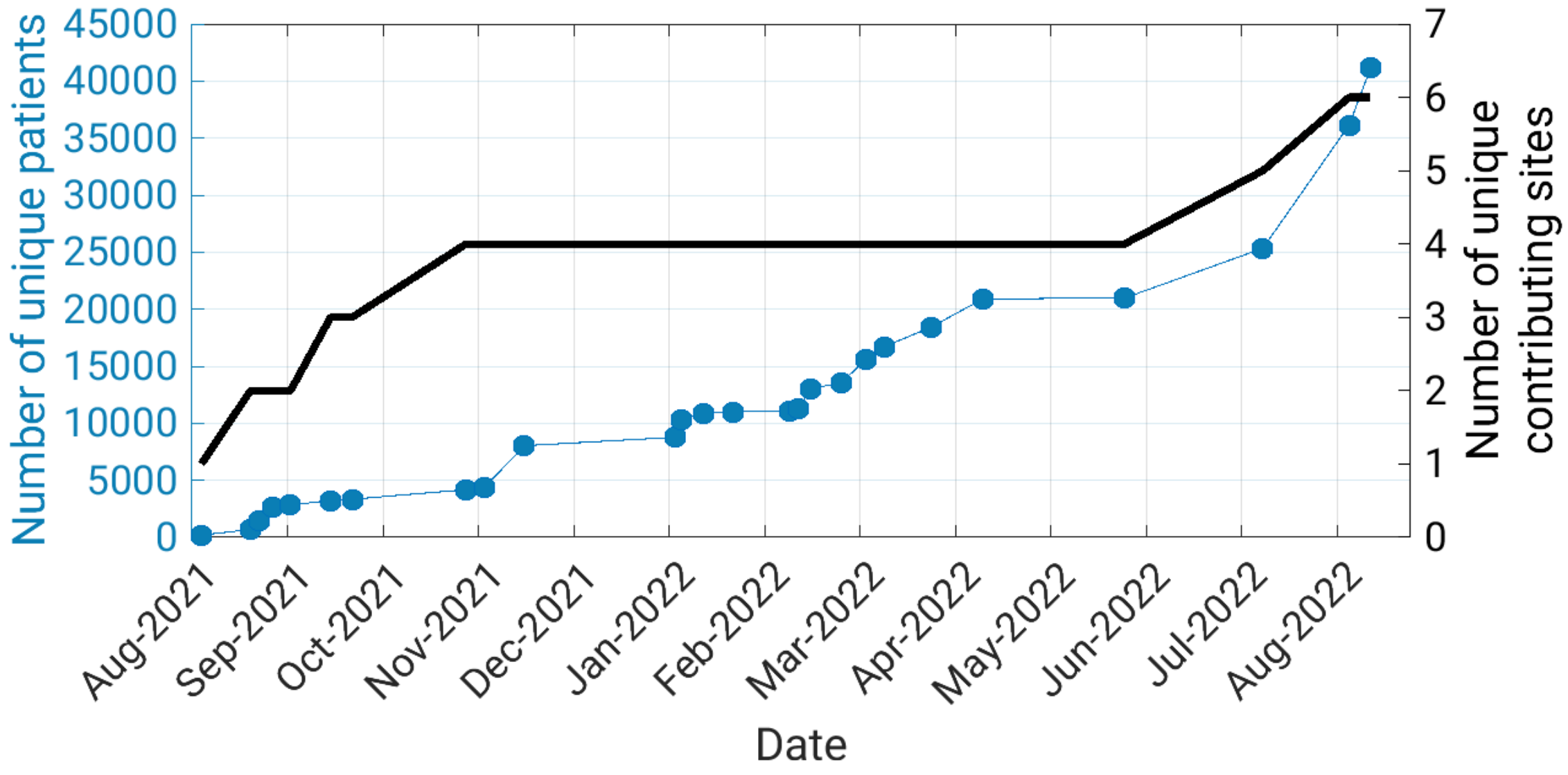

Figure 1: Cumulative number of unique patients and number of unique contributing sites in the open MIDRC data commons in the first year since the launch of the data commons in August 2021.

The proportions of the MIDRC data by demographic category and COVID-19 status as of August 19, 2022 are given in Figure 2.

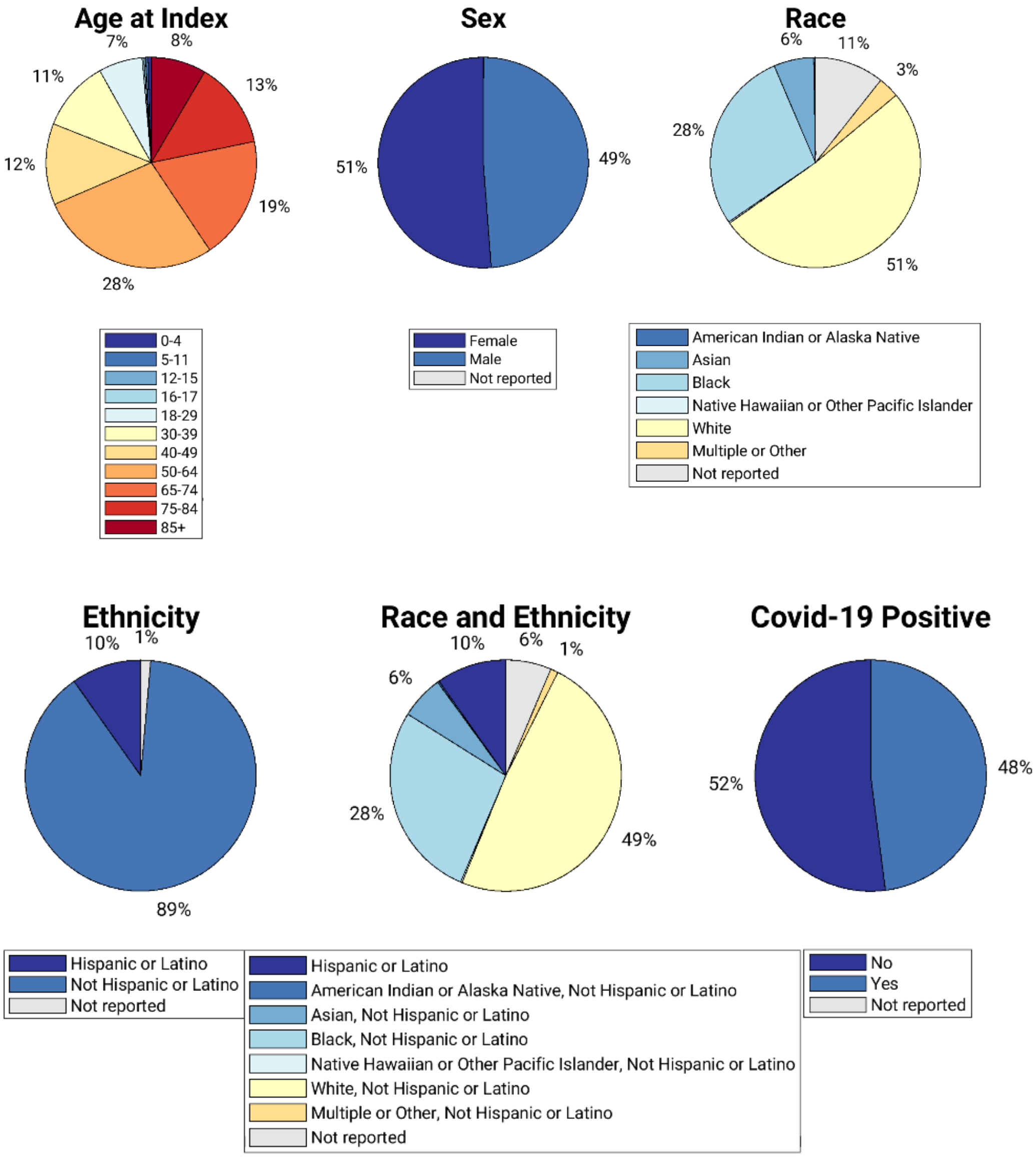

Figure 2: Pie charts of the percentages of unique patients in the MIDRC data as of August 19, 2022 by demographic category and COVID-19 status.

The most recent distributions of unique patients within each demographic category, both within the MIDRC data and the comparison groups, are given in Figure 3 along with JSD results. Longitudinal measurements of the demographic data for the MIDRC data are available in the Supplementary Data.

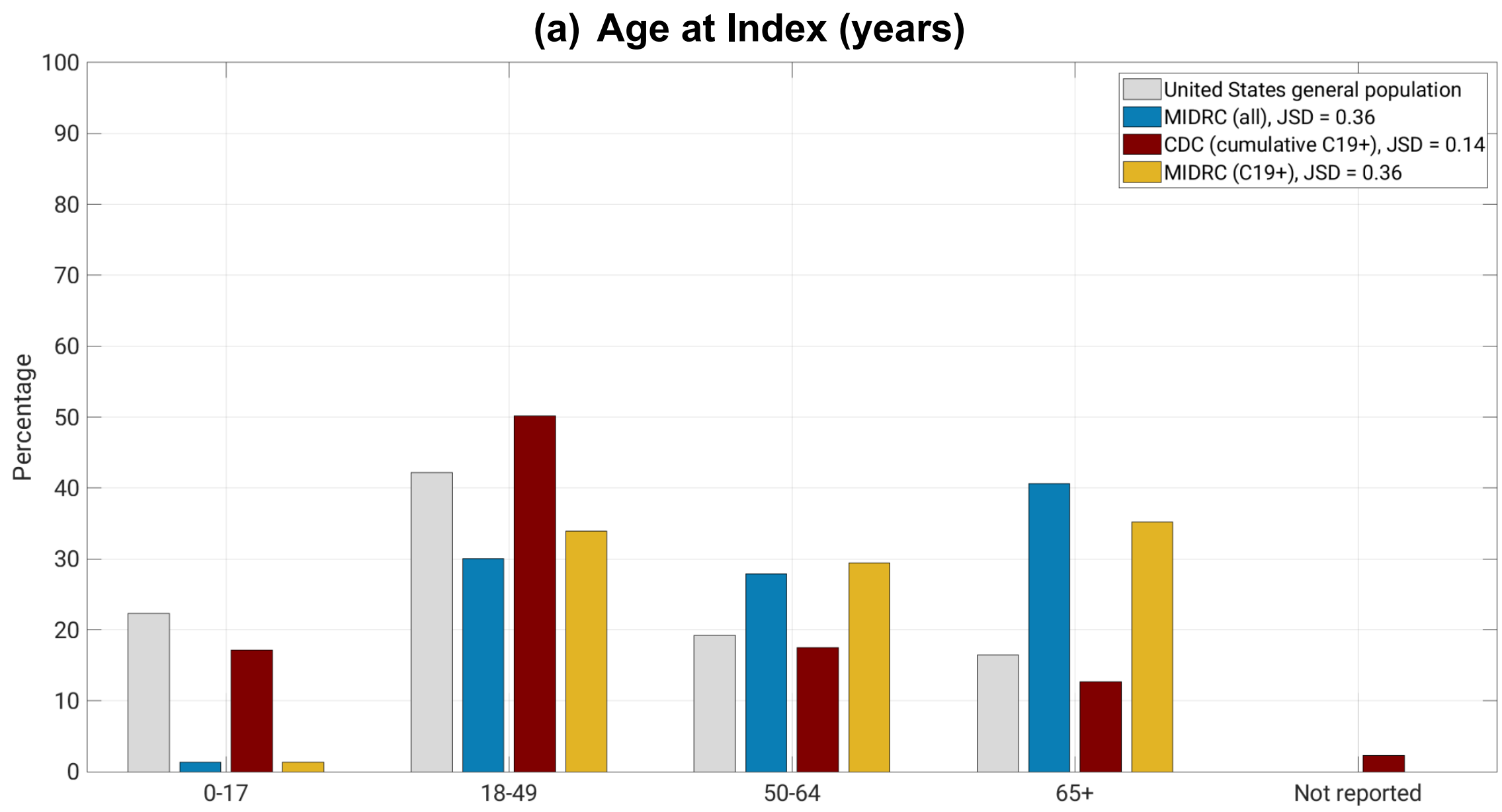
(a) Age at Index (years)
United States general population
MIDRC (all), JSD = 0.36
CDC (cumulative C19+), JSD = 0.14
MIDRC (C19+), JSD = 0.36
Percentage
100
90
80
70
60
50
40
30
20
10
0
0-17
18-49
50-64
65+
Not reported

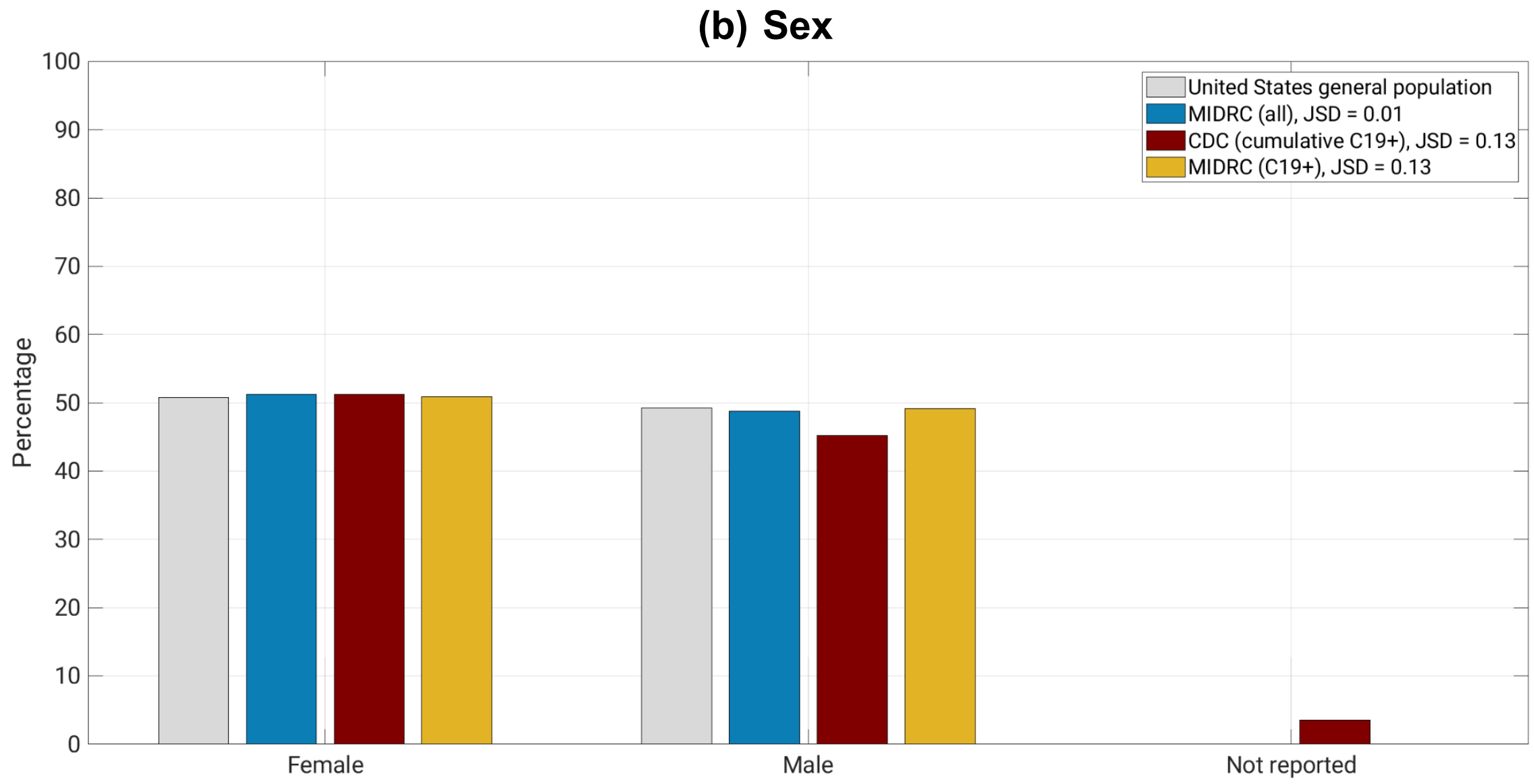

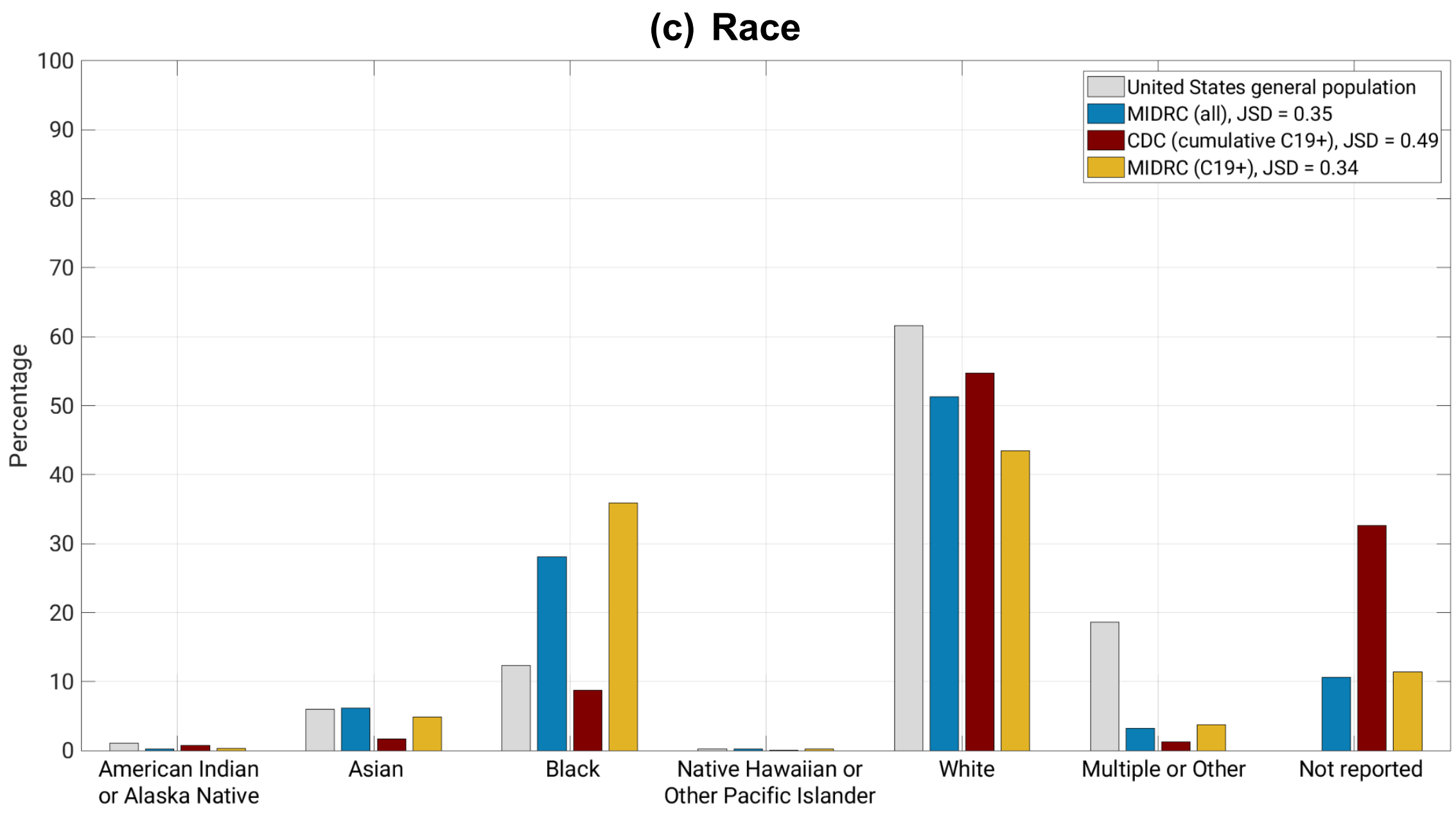
(c) Race
United States general population
MIDRC (all), JSD = 0.35
CDC (cumulative C19+), JSD = 0.49
MIDRC (C19+), JSD = 0.34
Percentage
100
90
80
70
60
50
40
30
20
10
0
American Indian or Alaska Native
Asian
Black
Native Hawaiian or Other Pacific Islander
White
Multiple or Other
Not reported

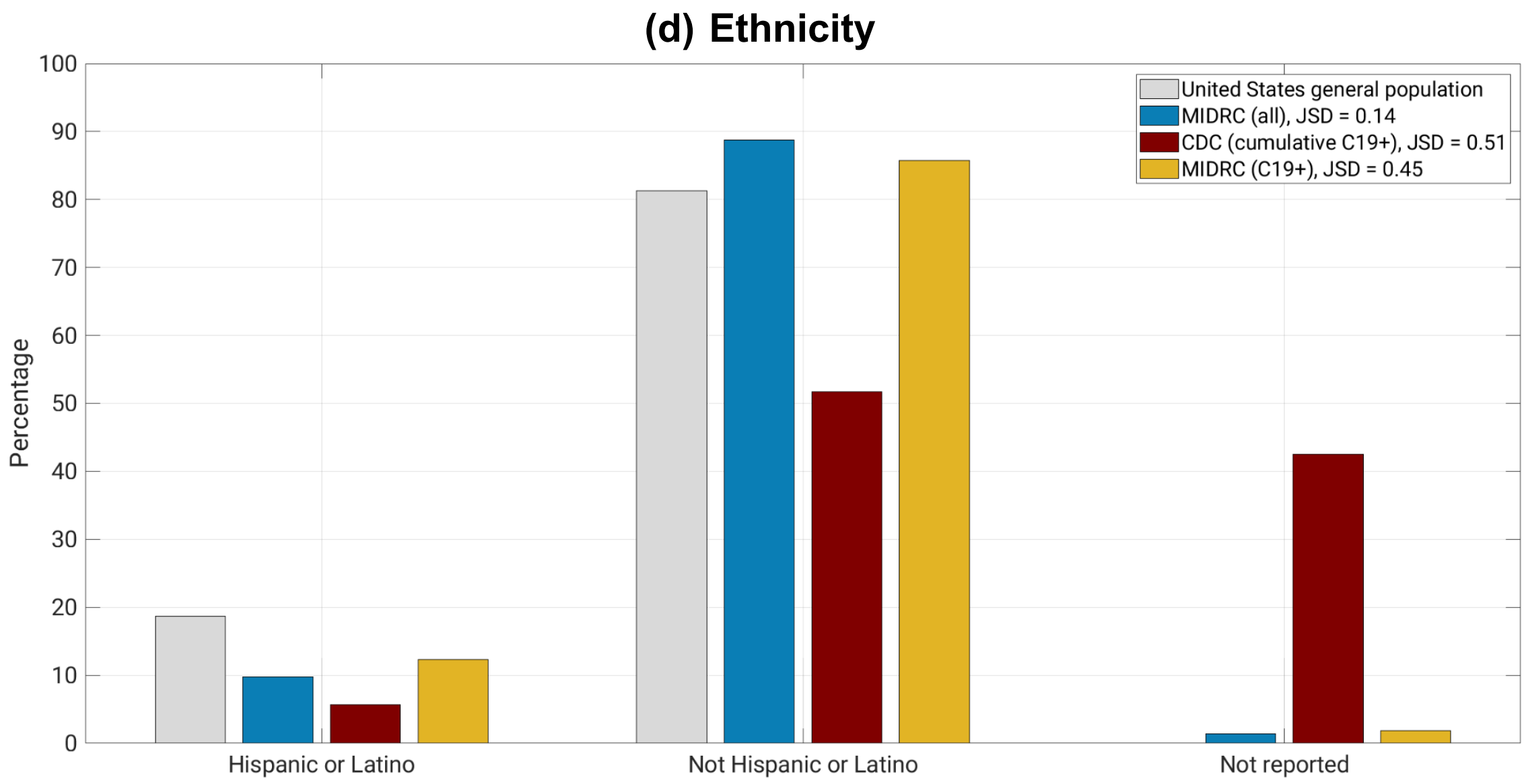
(d) Ethnicity
United States general population
MIDRC (all), JSD = 0.14
CDC (cumulative C19+), JSD = 0.51
MIDRC (C19+), JSD = 0.45
Percentage
100
90
80
70
60
50
40
30
20
10
0
Hispanic or Latino
Not Hispanic or Latino
Not reported

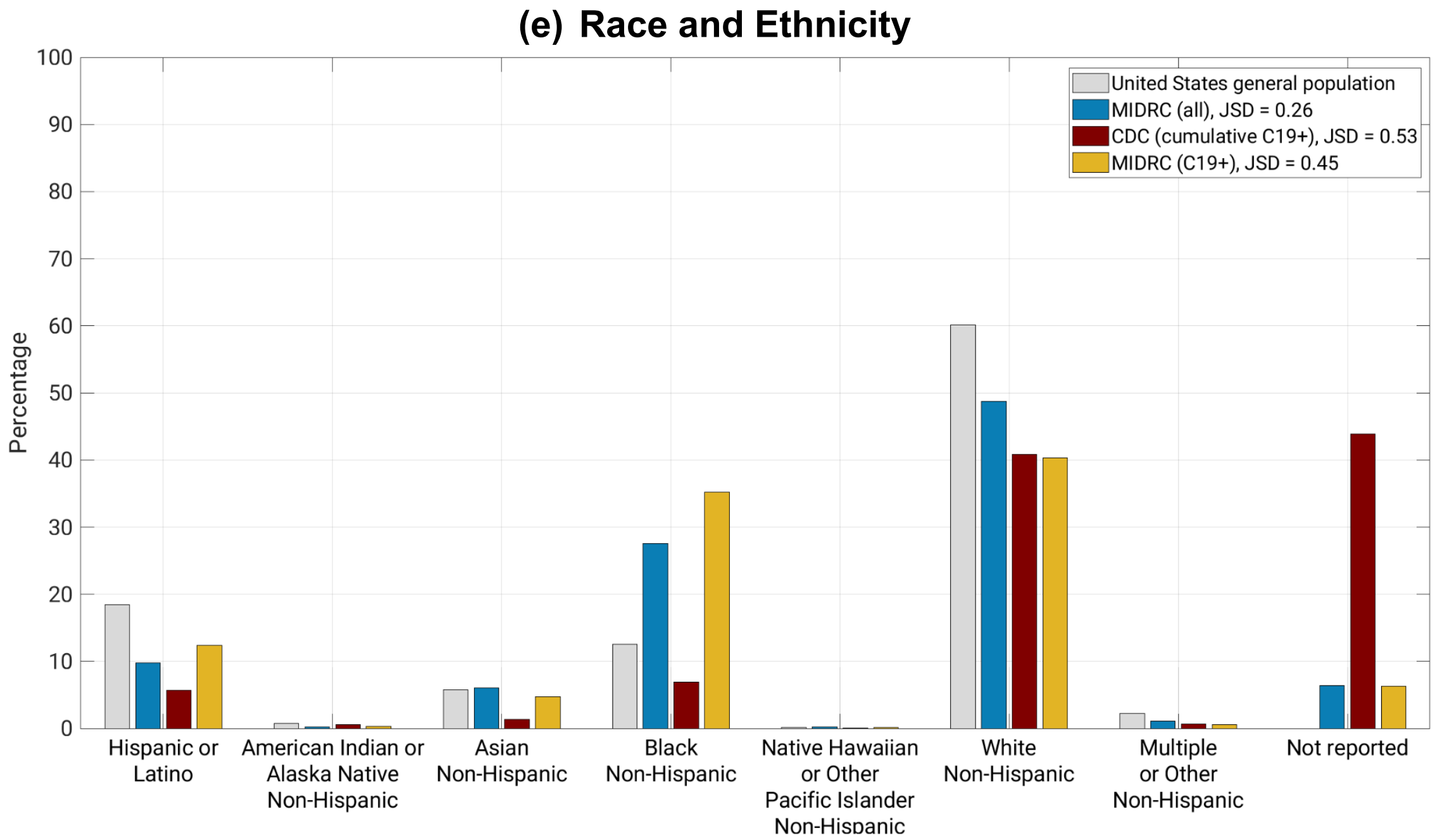

Figure 3: Distributions of cases within the MIDRC data as of August 19, 2022 and comparison groups US general population from the 2020 Census and cumulative case counts from the CDC. In these figures, the JSD is shown for (1) all cases in the MIDRC data compared to the US general population ($JSD_{\text{MIDRC (all) to Census}}$), (2) cumulative CDC COVID-19 positive case counts compared to the US general population ($JSD_{\text{CDC (C19+) to Census}}$), and (3) MIDRC COVID-19 positive case counts compared to the CDC COVID-19 positive case counts ($JSD_{\text{MIDRC (C19+) to CDC (C19+)}}$). The JSD is bounded between 0 to 1, where 0 indicates that two distributions are the same as measured by the JSD and 1 indicates that they are completely different. MIDRC: The Medical Imaging and Data Resource Center; US: United States; CDC: Centers for Disease Control and Prevention; C19+: cases positive for COVID-19.

The JSD measured changes in the similarity of the MIDRC data (both when considering all imaging studies and those from COVID-19 positive patients only) to their comparison groups (the United States general population and case counts from the CDC, respectively) (Figures 4-8). The comparison of age at index to the United States general population and the cumulative case counts as recorded by the CDC has remained relatively stable in these sets of patients over time, with little difference in their level of representativeness (Figure 4).

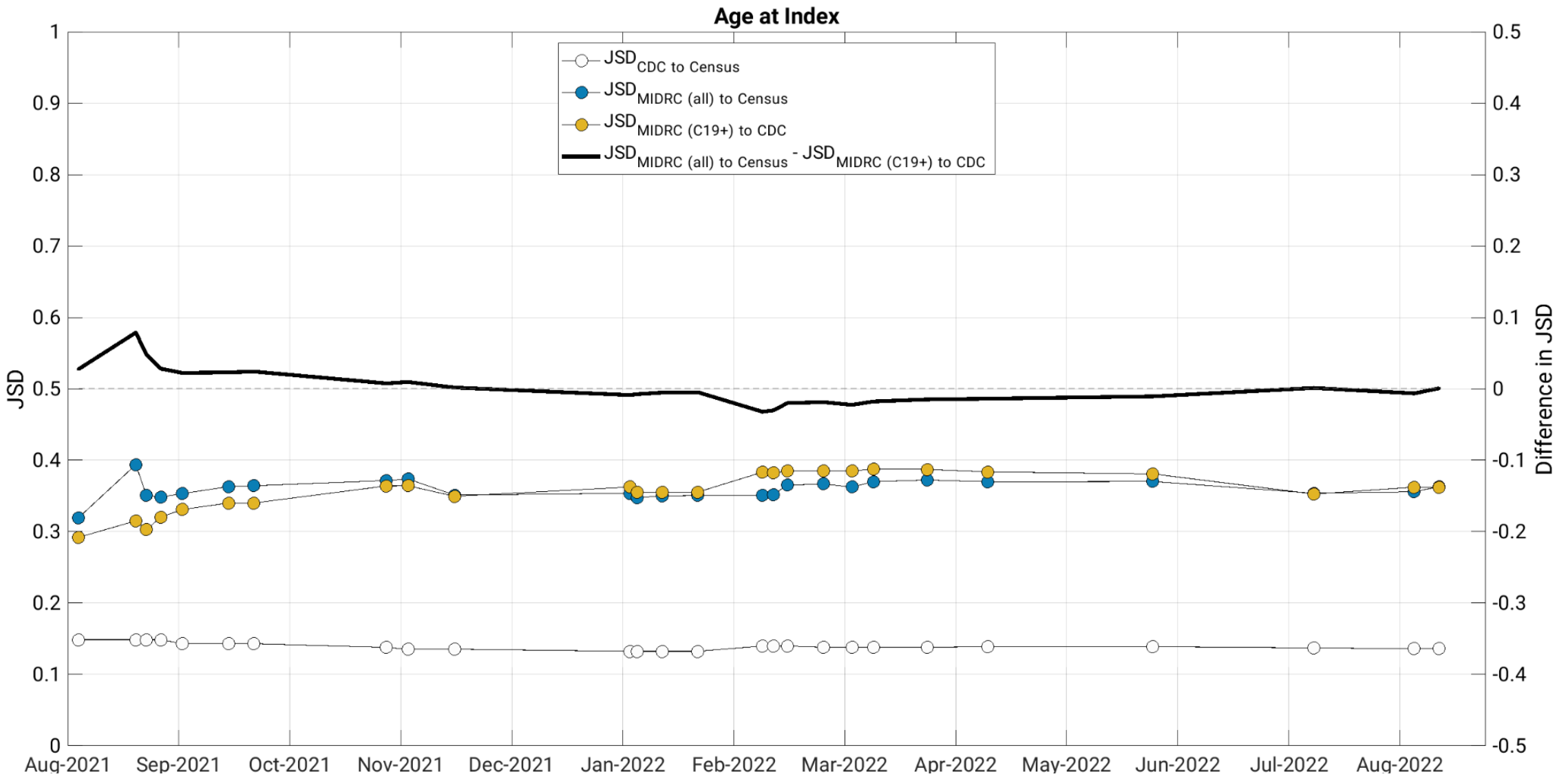

Figure 4: The Jensen-Shannon distance (JSD) over time for age at index for (blue data markers) all unique patients, (gold data markers) all unique COVID-19 positive patients in the MIDRC data, and (white data markers) the JSD for comparing the CDC data to the US general population (for reference). The difference in JSD over time between all unique patients and all unique COVID-19 positive patients in the MIDRC data is also shown (black line). The similarity of both all unique patients and all unique COVID-19 positive patients in the MIDRC data has remained fairly constant to their respective comparison groups over time in terms of JSD. MIDRC: The Medical Imaging and Data Resource Center; US: United States; CDC: Centers for Disease Control and Prevention; C19+: cases positive for COVID-19.

The comparison of MIDRC unique patients by sex to the United States general population and the cumulative case counts as recorded by the CDC has demonstrated more representativeness (i.e., more similarity) in the distribution of all unique patients to the United States general population (lower JSD) than the comparison of MIDRC positive patients to the cumulative case count from the CDC (higher JSD) (Figure 5). Over time, there has been a slight increase in the difference of the comparisons as the proportion of male unique patients has increased (Supplementary Data Figure 2).

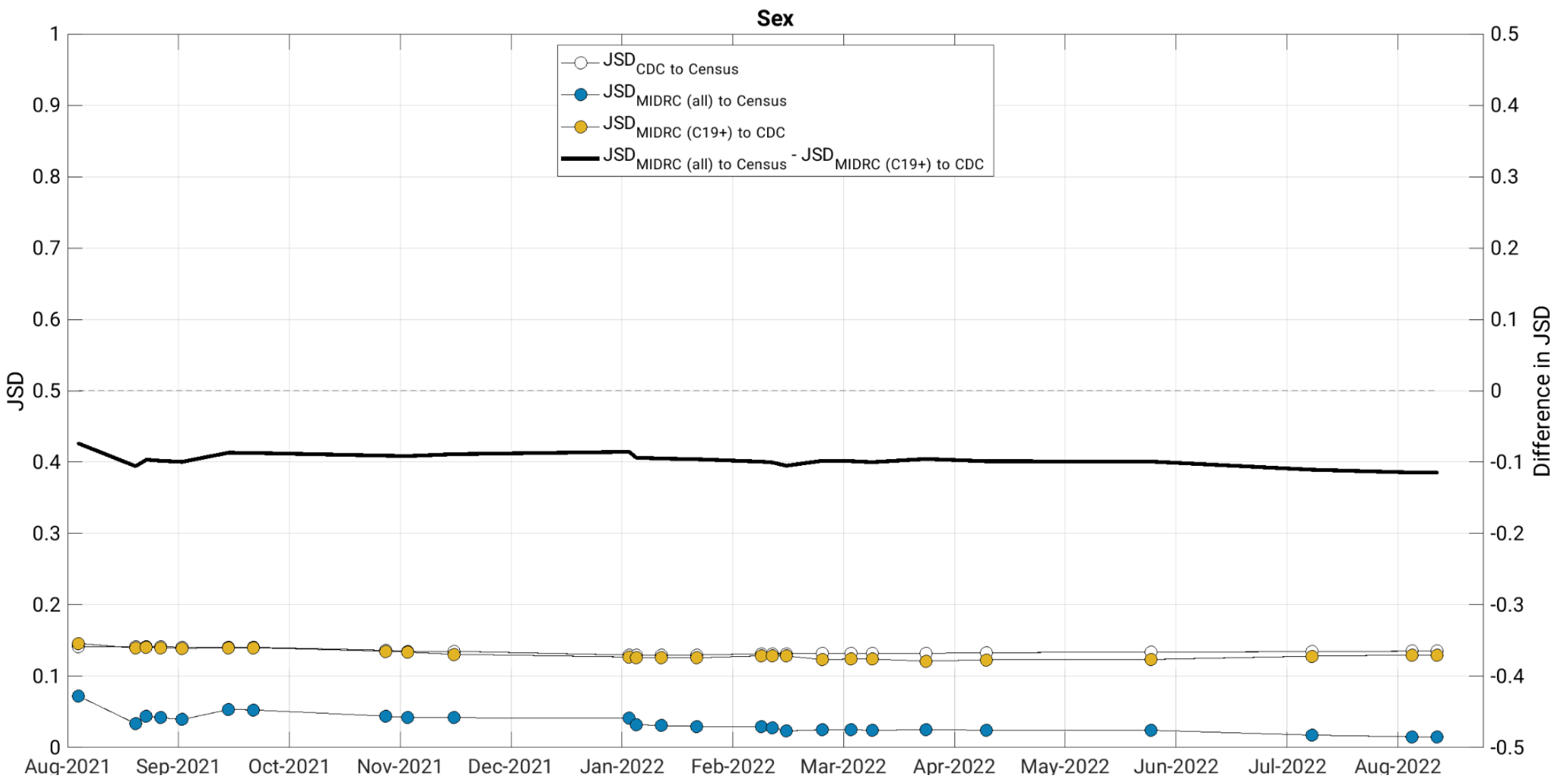

Figure 5: The Jensen-Shannon distance (JSD) over time for sex for (blue data markers) all unique patients, (gold data markers) all unique COVID-19 positive patients in the MIDRC data, and (white data markers) the JSD for comparing the CDC data to the US general population (for reference). The difference in JSD over time between all unique patients and all unique COVID-19 positive patients in the MIDRC data is also shown (black line). The distribution of all unique cases in the MIDRC data has been more representative of the US population than the distribution of all unique COVID-19 cases to the CDC cumulative case counts, and this higher representativeness has slightly increased as the number of unique cases has increased. MIDRC: The Medical Imaging and Data Resource Center; US: United States; CDC: Centers for Disease Control and Prevention; C19+: cases positive for COVID-19.

The representativeness of MIDRC unique patients by race to the United States general population and the cumulative case counts as recorded by the CDC has recently reached almost equal similarity within the MIDRC data (Figure 6). However, it is important to note that the measurement of representativeness for all three comparisons is impacted by the proportion of subjects for which race is not reported, which is over 30% in the most recent CDC cumulative case counts and around 10% in the most recent distributions within the MIDRC data. The MIDRC data also has substantially higher proportions of unique patients with reported race as Black than the US general population and the CDC cumulative case counts.

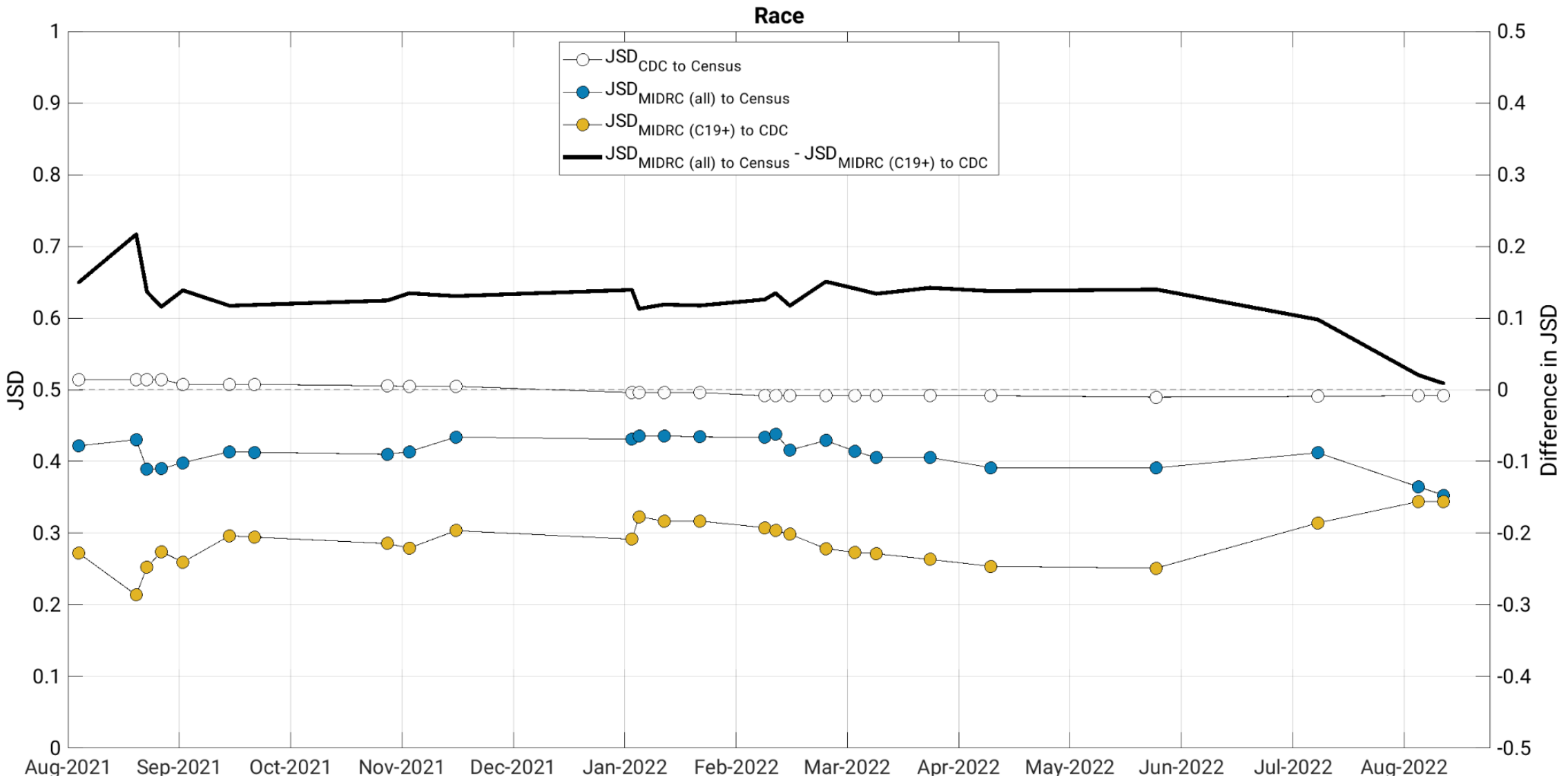

Figure 6: The Jensen-Shannon distance (JSD) over time for race for (blue data markers) all unique patients, (gold data markers) all unique COVID-19 positive patients in the MIDRC data, and (white data markers) the JSD for comparing the CDC data to the US general population (for reference). The difference in JSD over time between all unique patients and all unique COVID-19 positive patients in the MIDRC data is also shown (black line). The distribution of unique patients in the MIDRC data has recently reached similar levels of representativeness to their comparison groups (difference in JSD approaching zero). MIDRC: The Medical Imaging and Data Resource Center; US: United States; CDC: Centers for Disease Control and Prevention; C19+: cases positive for COVID-19.

The comparison of MIDRC unique patients by ethnicity to the United States general population has been substantially more similar than the comparison of MIDRC COVID-19 positive patients to the cumulative CDC case counts (Figure 7). This likely a result of the substantial percentage of cases within the CDC data for which ethnicity is not available (over 40%).

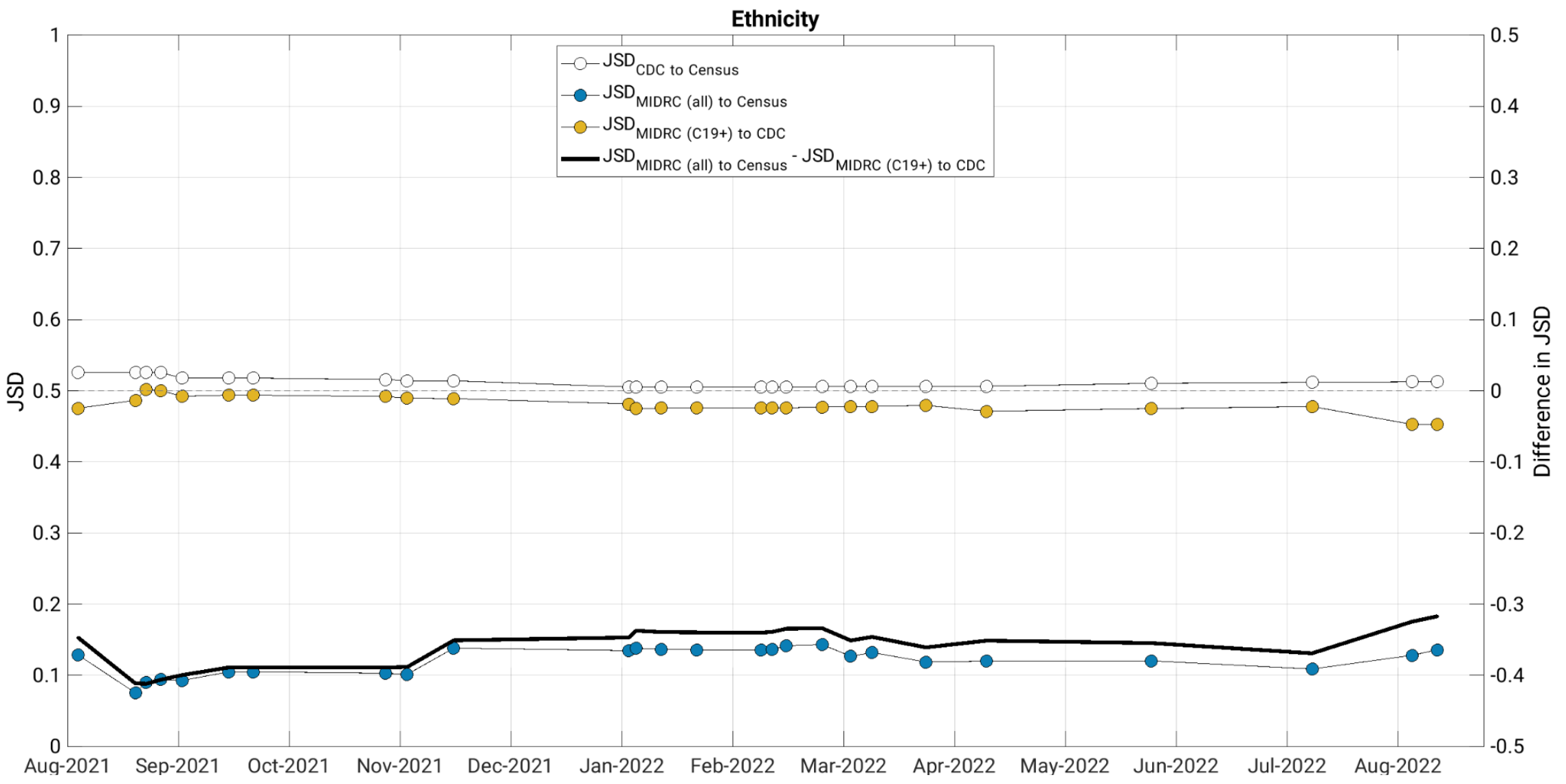

Figure 7: The Jensen-Shannon distance (JSD) over time for ethnicity for (blue data markers) all unique patients, (gold data markers) all unique COVID-19 positive patients in the MIDRC data, and (white data markers) the JSD for comparing the CDC data to the US general population (for reference). The difference in JSD over time between all unique patients and all unique COVID-19 positive patients in the MIDRC data is also shown (black line). The distribution comparisons are substantially similar for all unique patients in MIDRC to the United States general population than all unique COVID-19 positive patients to the case count distributions from the Centers for Disease Control and Prevention (CDC) due to the high percentage of cases within the CDC data for which ethnicity is not available (over 40%). MIDRC: The Medical Imaging and Data Resource Center; US: United States; CDC: Centers for Disease Control and Prevention; C19+: cases positive for COVID-19.

The comparison of MIDRC unique patients by the intersection of race and ethnicity to the United States general population has been more similar than the comparison of MIDRC COVID-19 positive patients to the cumulative CDC case counts (Figure 8). This may also be impacted by the substantial percentage of cases within the CDC data for which race and ethnicity is not available.

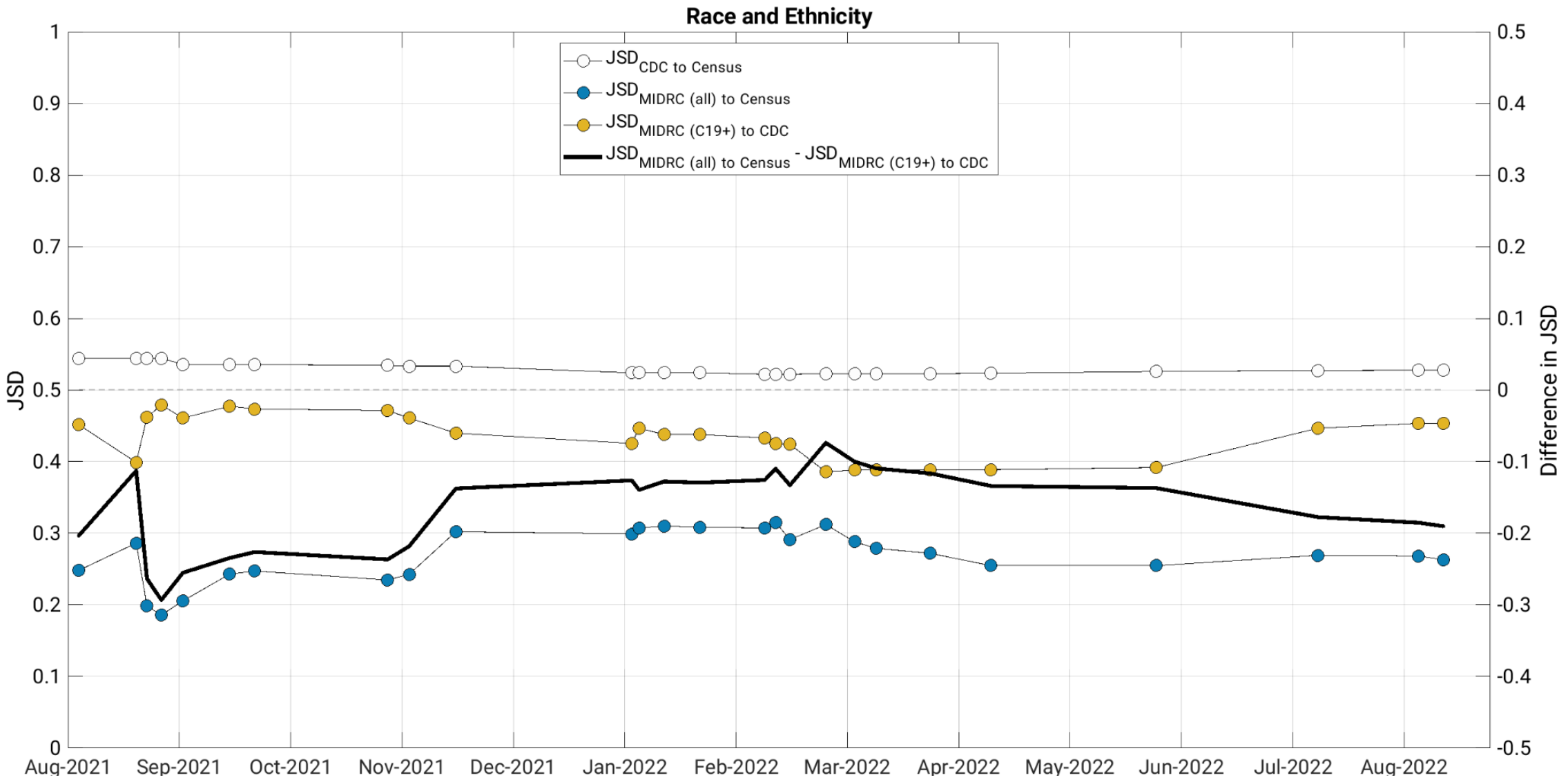

Figure 8: The Jensen-Shannon distance (JSD) over time for race and ethnicity for (blue data markers) all unique patients, (gold data markers) all unique COVID-19 positive patients in the MIDRC data, and (white data markers) the JSD for comparing the CDC data to the US general population (for reference). The difference in JSD over time between all unique patients and all unique COVID-19 positive patients in the MIDRC data is also shown (black line). The lower representativeness (higher JSD) of all unique COVID-19 positive patients in the MIDRC data to the CDC data and the CDC data to the US general population is impacted by the substantial percentage of cases within the data from the Centers for Disease Control and Prevention for which race and ethnicity is not reported (over 40%). MIDRC: The Medical Imaging and Data Resource Center; US: United States; CDC: Centers for Disease Control and Prevention; C19+: cases positive for COVID-19.

## Discussion

The representativeness of the MIDRC data continues to change over time as both the number of contributing institutions and the overall number of unique patients grow. The evolution of the impact of the COVID-19 pandemic to various demographic groups also continues to change over time, as shown by the changes in $JSD_{CDC\ (C19+)\ to\ Census}$ (not discussed in detail here). Using the JSD contributes to quantifying the comprehensive representativeness of the data and supports several initiatives related to health-related research and development at the federal level, including the strategic plan of the National

Institute on Minority Health and Health Disparities[30] (specifically, Goal 7: "Ensure appropriate representation of minority and other health disparity populations in NIH-funded research") and Action Plans and Guidance from the Food and Drug Administration.[31–33]

The goal of the development of fair and generalizable AI/ML algorithms in medical imaging has rightfully been the topic of much attention.[34–40] Goals of algorithmic fairness and generalizability involve both those of equal outcomes (including equalized outcomes) and of equal performance.[41] It should be noted that defining the representativeness of data is a crucial part of developing and deploying algorithms with fairness and generalizability in mind. Indeed, defining representativeness is one of many careful procedures needed in AI/ML, with others including but not limited to definition of the purpose of data collection, aim of the model, and careful identification of the task. As has been noted by others,[42] representativeness can involve representativity of data in the sense of coverage of the "input space" (i.e., the training and/or the test data) and/or representativeness to population distributions. The use of the JSD generally can support the assessment of representativeness of distributions for either aim; in this study we used the JSD to measure representativeness of the MIDRC data to demographic distributions.

Assessment of representativeness of data by demographic categories is but one part of ensuring fairness and generalizability at various stages of AI/ML pipelines. These include data collection (by identifying protected groups and their representation and addressing unequal representation in data through intentional collection efforts), model tuning and evaluation (such as comparing deployment data with training data across subgroups), and performance monitoring (including monitoring for data shifts, such as changes of impact in disease to subgroups over time[41,54]). Population characterization (and potentially matching synthetically[43,44]), cross-population modeling,[45,46] and class balancing[47] can be useful in AI/ML algorithm development to identify and avoid model bias.

We believe the Jensen-Shannon distance to be useful in AI/ML investigations in medical imaging, in part due to its intuitive nature (especially in terms of its bounds) and its relationship to information theory, which is an important foundation to other AI/ML performance measures such as receiver operating characteristic analysis.[48] Jensen-Shannon measures have also been used in some biology studies.[49] There are other methods for comparing the distributions of populations, such as the Heillinger distance,[50] population matching discrepancy[51] and matching quantiles estimation.[52] The ratio of patient identity to disease prevalence, termed the participant to prevalence ratio (PPR),[53] is used in some clinical trials (in which subjects are termed "participants") to measure representation within demographic subgroups, including intersectional identity. A PPR from 0.8 to 1.2 is largely considered adequate representation, but the establishment of this range appears to be somewhat ad hoc. A measure such as the JSD is desirable for our definition of representativeness due to its ability to summarize across a demographic category, but measures such as the PPR would be complementary for analysis of individual subgroups. It would be advantageous for future studies to quantify the impact of ranges of PPR on adequate representation or lack thereof and to establish more specific criteria for such levels. On the whole, it will be useful to conduct a comprehensive comparison of different measures of representativeness (both across an entire demographic category and by subgroups) and their relationship to fairness of AI in medical imaging; this will be the topic of future work.

There were some limitations to this study. First, the demographic categories described here were limited to one intersectional identity (race and ethnicity). Other intersectional identities are important to consider (such as that of age and race or sex and race) and will be the topic of future study. Secondly, this study did not include other factors which may be relevant in studying health inequities, such as patient residence (e.g., urban versus rural), healthcare institution type (e.g., community versus academic), patient education level attainment, patient income or experienced income equities, and patient employment status. Thirdly, there were some limitations in the data reported by the CDC: (1) it includes both probable and lab-confirmed COVID-19 cases, while COVID-19 positive cases in the MIDRC data commons include lab confirmation, and (2) it includes non-unique case counts (i.e., the counts include some individuals who have tested positive for COVID-19 at different times) while the MIDRC data counts patients only once. Finally, MIDRC works with data contributors to receive imaging study donations which have been de-identified using the Safe Harbor method, in compliance with the Health Insurance Portability and Accountability Act of 1996 (HIPAA) Privacy Rule. Thus, while each patient's timeline is preserved, the MIDRC data commons does not provide the actual date of image acquisition and COVID test. This means that the imaging studies within a given ingestion date can include imaging studies acquired theoretically at any time *before* the ingestion date, necessitating our comparison of cumulative case counts in both the MIDRC data and the COVID-19 case counts from the CDC, rather than a potential comparison for cases imaged within a given month to cases reported by the CDC within a given month.

In summary, the demographic characteristics of the MIDRC data in the categories of age at imaging, sex, race, ethnicity, and the intersection of race and ethnicity and their similarity to comparison groups were measured using the Jensen-Shannon distance. Overall, the JSD indicated more representativeness for all unique patients than for COVID-19 positive patients when compared to their respective comparison groups. These measures can be used by investigators in developing unbiased and generalizable AI/ML algorithms using the MIDRC data, including when building cohorts.

**Acknowledgements**

The authors are grateful to the staff at Gen3, the staff of ACR, RSNA, and AAPM, and the data contributors. The Medical Imaging and Data Resource Center (MIDRC) is funded by the National Institute of Biomedical Imaging and Bioengineering (NIBIB), part of the National Institutes of Health under contracts 75N92020C00008 and 75N92020C00021.

**Conflicts of interest**

Dr. Kalpathy-Cramer has no funding to report for this manuscript but funding for other work unrelated to what is presented here includes a research grant from GE, research support from Genentech, Consultant/stock options from Siloam Vision, LLC and technology licensed to Boston AI.

Dr. Koyejo has received funding support through awards NSF IIS 2205329 and NSF IIS 2046795.

Dr. Myers works as an independent technical and regulatory consultant as principal for Puente Solutions LLC.

Dr. Wawira-Gichoya has received funding support from the US National Science Foundation #1928481 from the Division of Electrical, Communication & Cyber Systems.

## Author biographies

**Natalie Baughan** is a PhD candidate in the University of Chicago Graduate Program in Medical Physics. After receiving her BS degree in nuclear engineering and radiological sciences from the University of Michigan in 2019, her research has focused on breast cancer risk assessment in mammography and statistical methods for AI.

**Brad Bower, PhD,** contributes as a Data and Technology Advancement (DATA) National Service Scholar with the National Institutes of Health (NIH) Office of Data Science Strategy (ODSS) and the National Institute of Biomedical Imaging and Bioengineering. He has a background in developing and commercializing imaging and data-driven devices in healthcare. At NIBIB he is leading efforts for improving FAIR data use and bench-to-bedside AI healthcare product development.

**Weijie Chen, PhD**, is a Research Physicist in the Division of Imaging, Diagnostics, and Software Reliability, Office of Science and Engineering Laboratories, CDRH, US FDA, where he conducts research and provides consults to regulatory review of medical devices. He earned his PhD in Medical Physics in 2007 from the University of Chicago. His research interests include performance characterization and assessment methodologies for medical imaging and AI/ML/CAD devices.

**Karen Drukker, PhD**, is a research associate professor of Radiology at the University of Chicago, where she has been involved in medical imaging research for 20+ years. She received her PhD in physics from the University of Amsterdam. Her research interests include machine learning applications in the detection, diagnosis, and prognosis of disease, focusing on rigorous training/testing protocols, generalizability, performance evaluation, and bias and fairness of AI. She is a fellow of SPIE.

**Maryellen Giger, Ph.D.** is the A.N. Pritzker Distinguished Service Professor at the University of Chicago. Her research involves computer-aided diagnosis/machine learning in medical imaging for cancer and now COVID-19, and is contact PI on the NIBIB-funded Medical Imaging and Data Resource Center (MIDRC; midrc.org), which has published more than 100,000 medical imaging studies for use by AI investigators. She is a member of the NAE, a recipient of the AAPM Coolidge Gold Medal, SPIE Harrison H. Barrett Award, and RSNA Outstanding Researcher Award, and is a Fellow of AAPM, AIMBE, SPIE, IEEE.

**Nicholas Gruszauskas, PhD,** is currently the Technical Director of the University of Chicago's Human Imaging Research Office and a faculty member in the Department of Radiology's Clinical Imaging Medical Physics Residency Program. He earned both an M.S. and Ph.D. in Biomedical Engineering from the University of Illinois Chicago, where he investigated computer-aided diagnosis and artificial intelligence methods for breast imaging. His team is currently responsible for facilitating clinical trial medical imaging at the University.

**Jayashree Kalpathy-Cramer, PhD**, is the chief of the Division of Artificial Medical Intelligence in the Department of Ophthalmology at the University of Colorado. Previously, she was an Associate Professor of Radiology at Harvard Medical School where she was actively involved in data science activities with a focus on medical imaging. Her research spans the spectrum from novel algorithm development to clinical deployment. Dr. Kalpathy-Cramer has authored over 200 peer-reviewed publications and has written over a dozen book chapters.

**Sanmi Koyejo, PhD**, is an Assistant Professor in the Department of Computer Science at Stanford University. Koyejo's research interests are in developing the principles and practice of trustworthy machine learning, including fairness and robustness. Additionally, Koyejo focuses on applications to neuroscience and healthcare. Koyejo has received several awards, including a best paper award from the conference on uncertainty in artificial intelligence, a Skip Ellis Early Career Award, and a Sloan Fellowship.

**Kyle Myers, PhD**, served as a research scientist and manager in the FDA's Center for Devices and Radiological Health for over 30 years. She coauthored *Foundations of Image Science*, winner of the First Biennial J.W. Goodman Book Writing Award from OSA and SPIE. Dr. Myers is a Fellow of AIMBE, Optica, SPIE, and a member of the National Academy of Engineering. She received her Ph.D. in Optical Sciences from the University of Arizona in 1985.

**Rui Carlos Pereira de Sá, PhD**, is an Assistant Professor of Physiology at the University of California, San Diego. His research focuses on functional imaging of the human lung in health and disease. As a Data and Technology Advancement (DATA) National Service Scholar at NIBIB/ NIH, R.C. Sá supported the first two years of MIDRC and other NIH medical imaging data-centric initiatives.

**Berkman Sahiner, PhD,** is a senior biomedical research scientist with the Division of Imaging, Diagnostics and Software Reliability, Center for Devices and Radiological Health, U.S. FDA. He has a PhD in electrical engineering and computer science from the University of Michigan, Ann Arbor.His research is focused on the evaluation of medical imaging and computer-assisted diagnosis devices, including devices that incorporate machine learning and artificial intelligence. He is a fellow of SPIE and AIMBE.

**Judy Wawira-Gichoya**, PhD, is an assistant professor at Emory university in Interventional Radiology and Informatics. Her career focus is on validating machine learning models for health in real clinical settings, exploring explainability, fairness, and a specific focus on how algorithms fail. She is heavily invested in training the next generation of data scientists through multiple high school programs, serving as the program director for radiology:AI trainee editorial board and the medical students machine learning elective.

**Heather Whitney, PhD**, is a Research Assistant Professor of Radiology at the University of Chicago. Her experience in quantitative medical imaging has ranged from polymer gel dosimetry to radiation damping in nuclear magnetic resonance to radiomics. She is interested in investigating the effects of the physical basis of imaging on radiomics, the repeatability and robustness of radiomics, the development of methods for task-based distribution, and bias and diversity of medical imaging datasets.

**Zi Jill Zhang, MD**, is an Assistant Professor of Radiology at the Sidney Kimmel Medical College Thomas Jefferson University. She completed her Imaging Informatics Fellowship and Breast Imaging Fellowship at the University of Pennsylvania. She is a member of the Bias and Diversity Working Group at the Medical Imaging and Data Resource Center. She also serves as a committee member of the RSNA Asia/Oceania International Advisory Committee and the Society of Breast Imaging Social Media Committee.

**Supplemental data**

The longitudinal measurement of demographics within the MIDRC data has changed over time (Supplemental Figures 1-5).

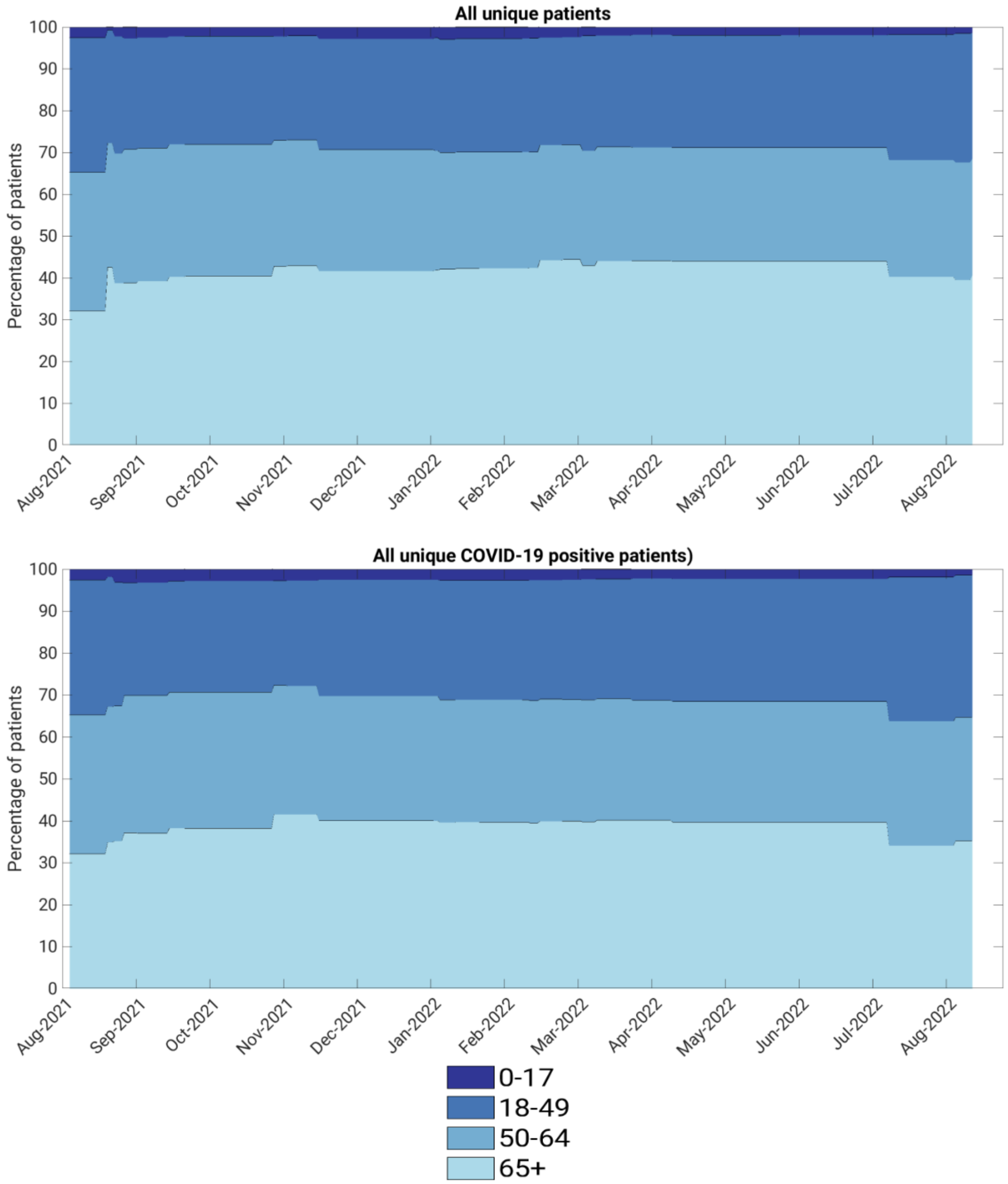

Supplemental Figure 1: Longitudinal percentage of patients in the demographic category of age (in years at index) for (top) all unique patients and (bottom) unique COVID-19 positive patients in the MIDRC data.

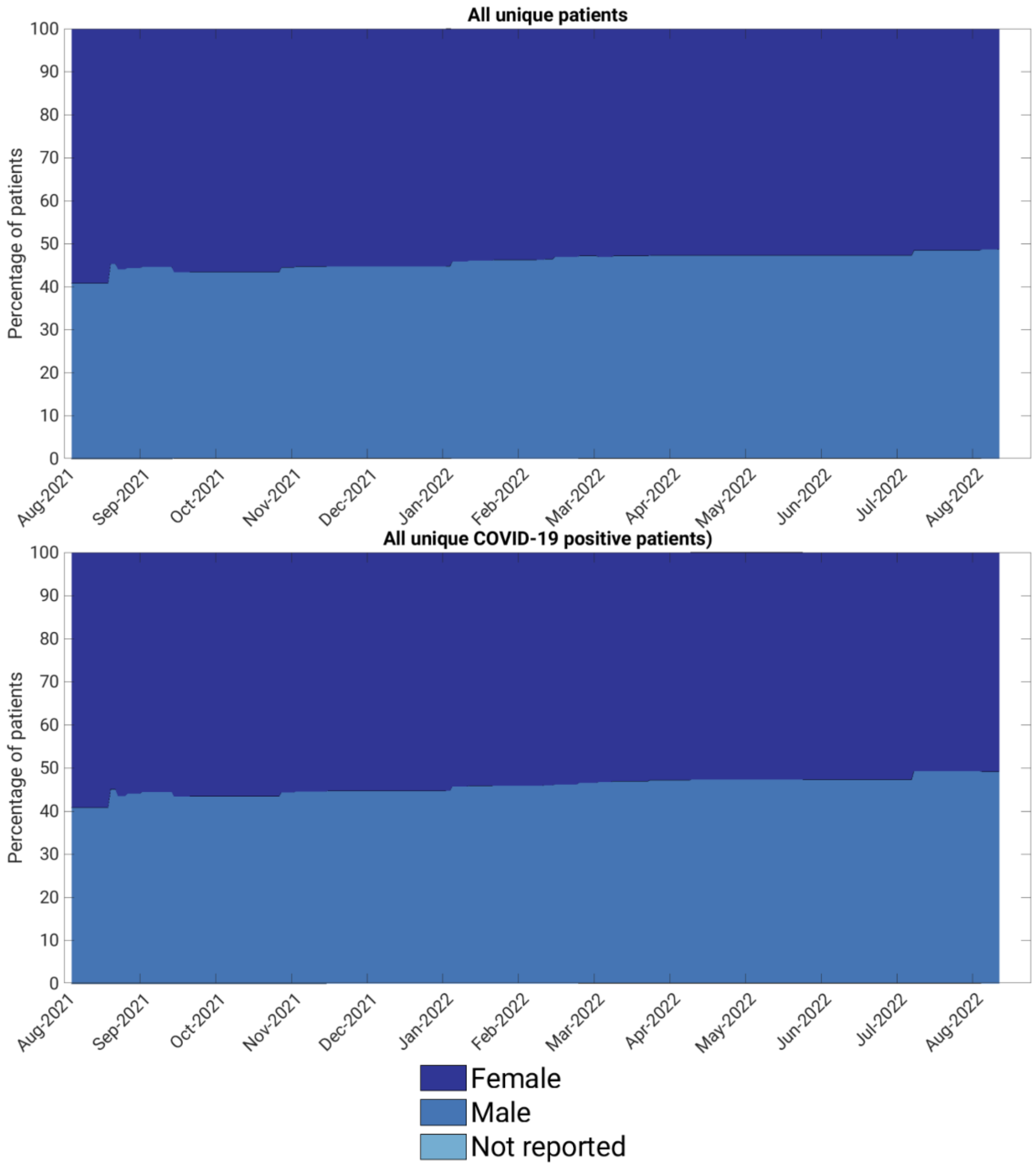

Supplemental Figure 2: Longitudinal percentage of patients in the demographic category of sex for (top) all unique patients and (bottom) unique COVID-19 positive patients in the MIDRC data.

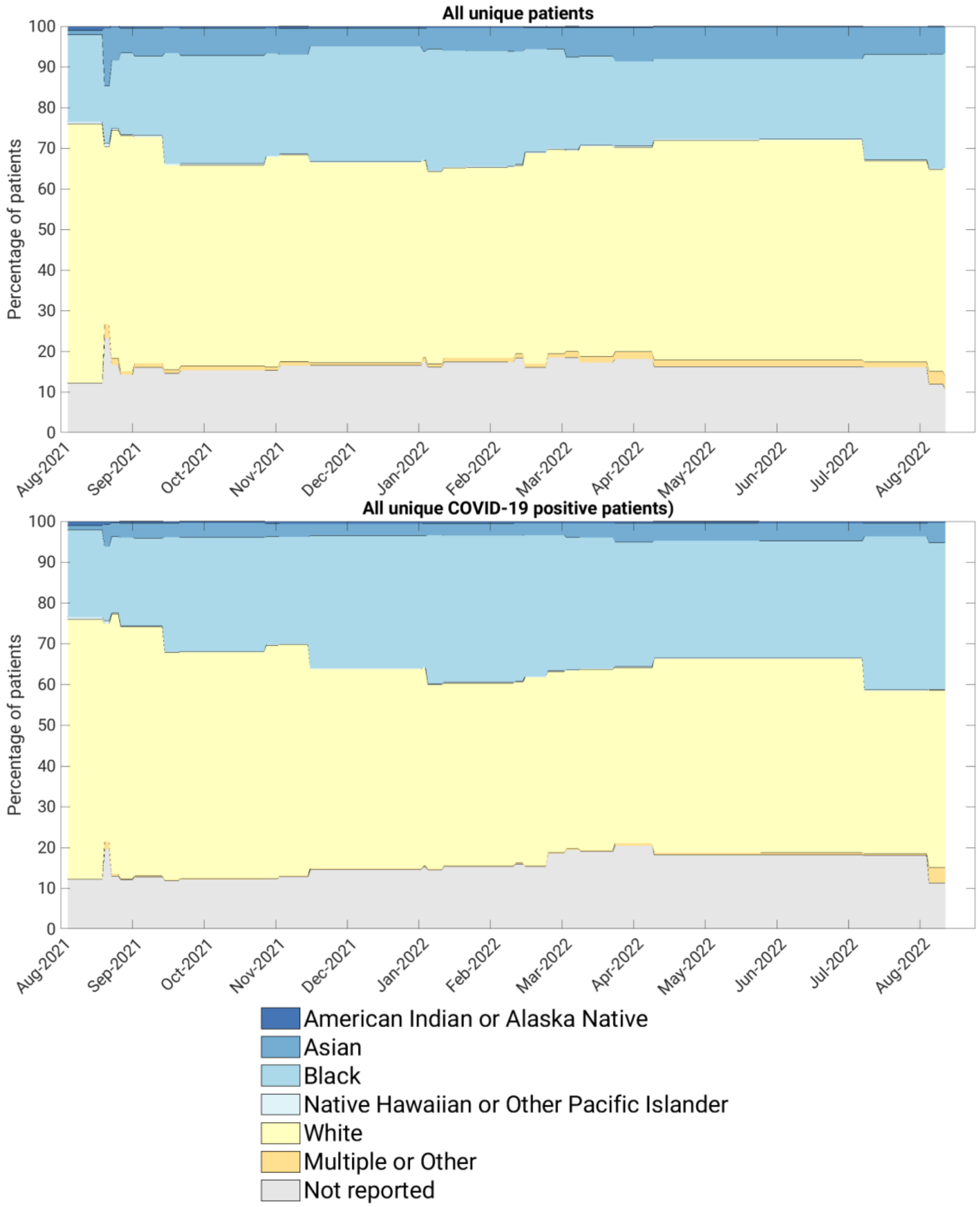

Supplemental Figure 3: Longitudinal percentage of patients in the demographic category of race for (top) all unique patients and (bottom) unique COVID-19 positive patients in the MIDRC data.

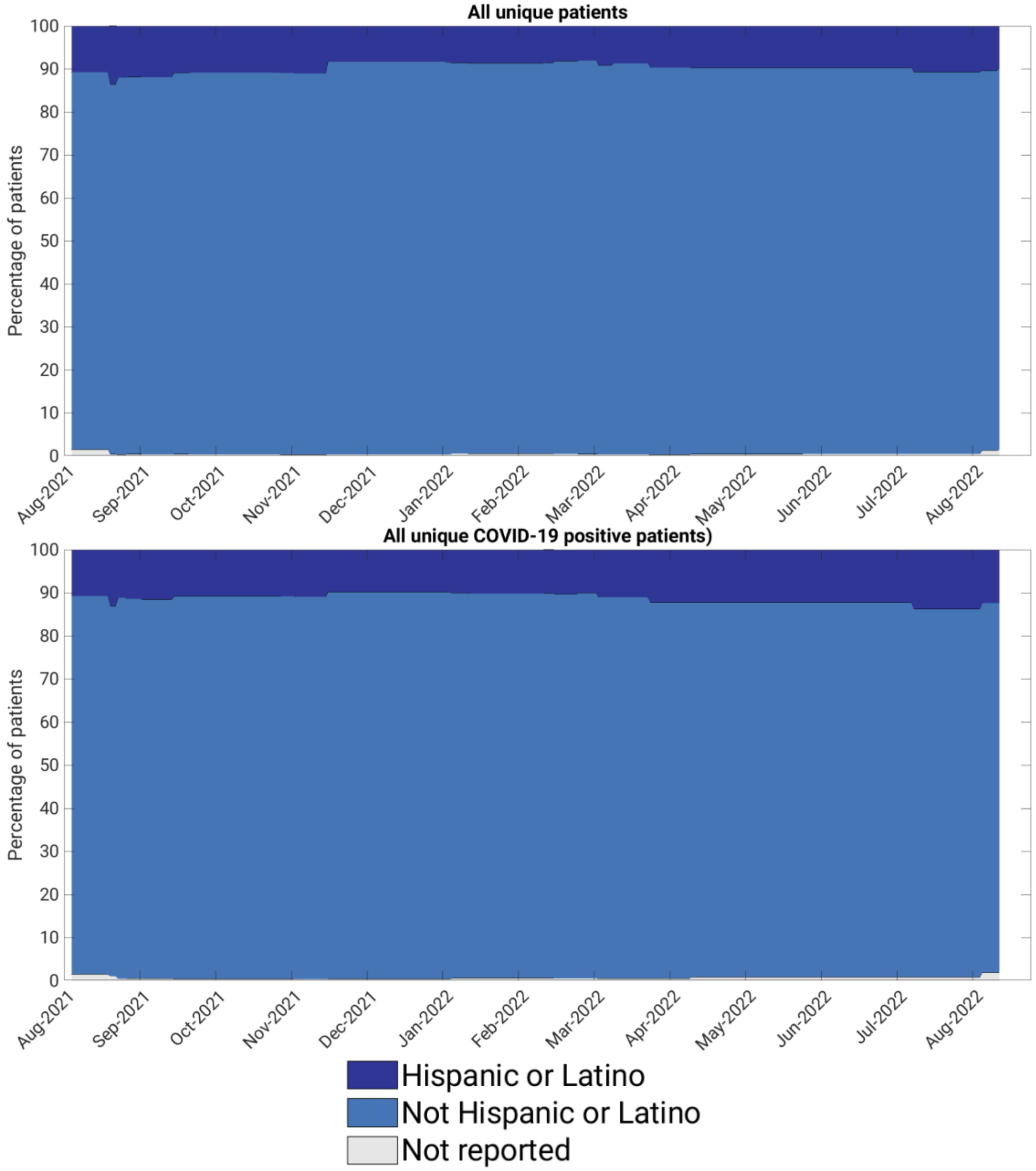

Supplemental Figure 4: Longitudinal percentage of patients in the demographic category of ethnicity for (top) all unique patients and (bottom) unique COVID-19 positive patients in the MIDRC data.

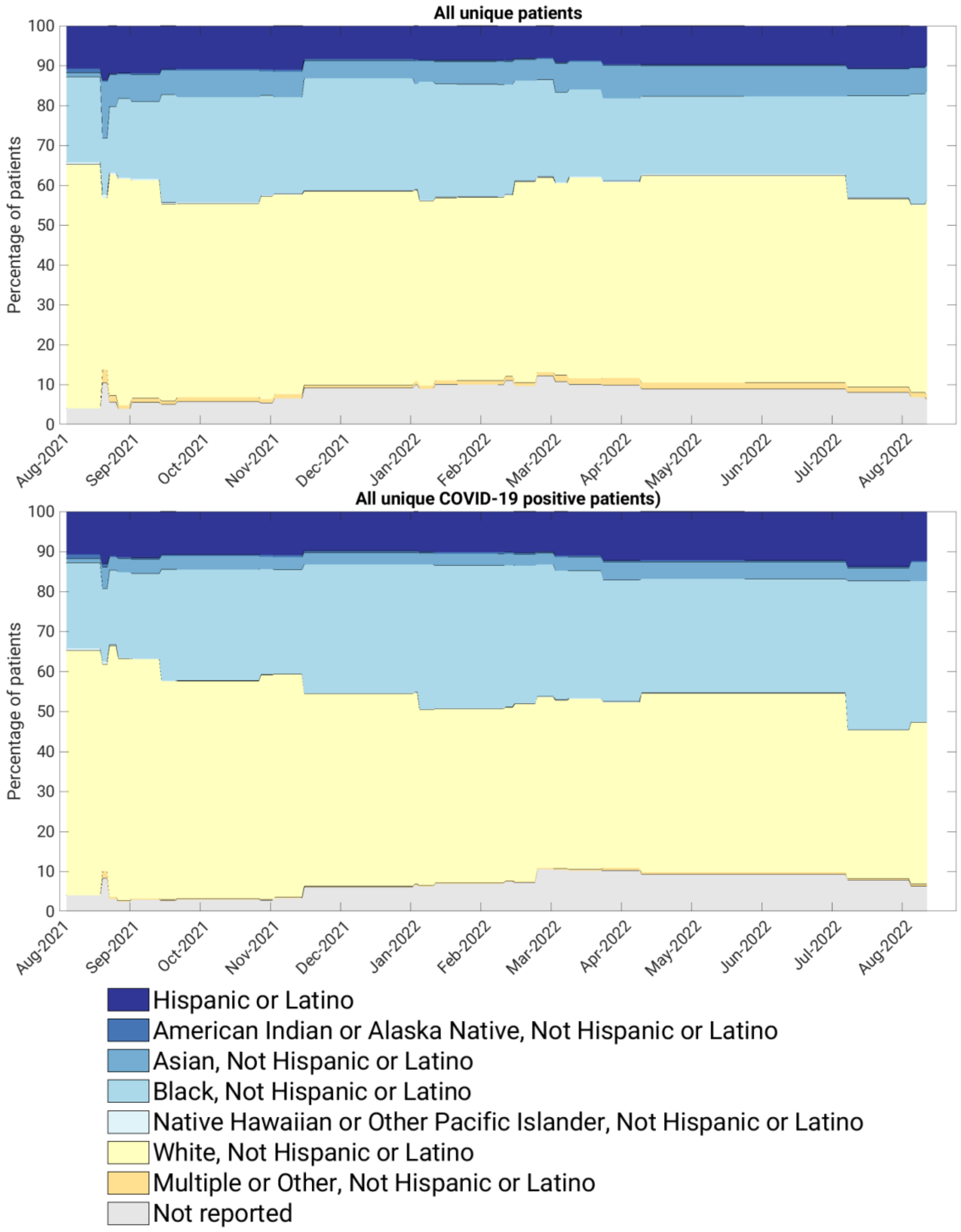

Supplemental Figure 5: Longitudinal percentage of patients in the demographic category of intersection of race and ethnicity for (top) all unique patients and (bottom) unique COVID-19 positive patients in the MIDRC data.